\documentstyle[aps,epsfig]{revtex}
\begin{document}
\draft
\tighten
\title{Dissipation, noise and DCC domain formation}
\author{\bf A. K. Chaudhuri \cite{byline}}
\address{ Variable Energy Cyclotron Centre\\
1/AF,Bidhan Nagar, Calcutta - 700 064\\}
\maketitle
\begin{abstract}
We  investigate  the  effect  of  friction on domain formation in
disoriented chiral condensate. We solve the equation of motion of
the linear sigma model, in the Hartree approximation, including a
friction and a white noise term. For quenched initial  condition,
we  find  that  even  in presence of noise and dissipation domain
like structure emerges after a few  fermi  of  evolution.  Domain
size as large as 5 fm can be formed. \end{abstract}


The possibility of forming disoriented chiral condensate (DCC) in
relativistic  heavy  ion  collisions  has  generated considerable
research activities in recent years. The idea was first  proposed
by Rajagopal and Wilczek \cite{ra93,ra93a,ra95,wi93}. They argued
that  for  a  second  order  chiral  phase transition, the chiral
condensate   can   become   temporarily   disoriented   in    the
nonequilibrium conditions encountered in heavy ion collisions. As
the  temperature drops below $T_c$, the chiral symmetry begins to
break  by  developing  domains  in  which  the  chiral  field  is
misaligned  from its true vacuum value. The misaligned condensate
has the same quark content and quantum numbers as  do  pions  and
essentially  constitute  a  classical pion field. The system will
finally relaxes to the true vacuum and in the  process  can  emit
coherent  pions.  Since the disoriented domains have well defined
isospin orientation,  the  associated  pions  can  exhibit  novel
centauro-like  \cite{la80,ar83,al86,re88} fluctuations of neutral
and charged pions \cite{an89,an89a,bj93,bl92}.

Most dynamical studies of DCC have been based on the linear sigma
model,  in  which  the chiral degrees of freedom are described by
the real O(4)  field  $\Phi=(\sigma,\roarrow{\Pi})$,  having  the
equation of motion,

\begin{equation}
[\Box + \lambda (\Phi^2-v^2)]\Phi=H n_\sigma
\label{1}
\end{equation}

The  parameters  of the model can be fixed by specifying the pion
decay constant, $f_\pi$=92 MeV and the meson masses,  $m_\pi$=135
MeV       and       $m_\sigma$=600      MeV,      leading      to
$\lambda=(m_\sigma^2-m_\pi^2)/2f_\pi^2$=20.14                 and
$v=[(m_\sigma^2-3            m_\pi^2)/(m_\sigma^2-m_\pi^2)]^{1/2}
f_\pi$=86.71  MeV  and  $H=(120.55  MeV)^3$  \cite{ra96}.  It  is
apparent  from  eq.\ref{1}  that  the  vacuum  is  aligned in the
$\sigma$  direction  $\Phi_{vac}=(f_\pi,\bf{0})$   and   at   low
temperature  the  fluctuations represent nearly free $\sigma$ and
$\pi$ mesons. At very high temperature well above $v$, the  field
fluctuations are centered near zero and approximate O(4) symmetry
prevails.

It is instructive to decompose the chiral field,

\begin{equation}
\Phi(r,t)=<\phi(r,t)>+\delta\phi(r,t) \label{2}
\end{equation}

\noindent  where $<\phi> $is the mean field and $\delta \phi$ are
the  semiclassical  fluctuations  around  $<\phi>$  and  can   be
identified  with quasi-particle excitations. Using eq.\ref{2} and
taking the average of eq.\ref{1}, the equation of motion for  the
mean fields in the  Hartree  approximation  can  be  obtained  as
\cite{ra96,as94},

\begin{equation}
\frac{\partial^2 <\phi>}{\partial t^2} -\nabla^2 <\phi>
= \lambda(v^2-<\phi>^2-
3<\delta  \phi^2_\|>-<\delta  \phi^2_\bot>)  <\phi>  +H  n_\sigma
\label{3}
\end{equation}

\noindent  where $<\phi>^2=<\phi_i><\phi_i>$, $\delta \phi_\|$ is
the component of the fluctuation parallel to $<\phi>$ and $\delta
\phi_\bot$ is the orthogonal component. This equation imply  that
the  motion  of  the  mean  field  is determined by the effective
potential,

\begin{equation}
V(<\phi>)=\frac{\lambda}{4}
(<\phi>^2+ 3<\delta \phi^2_\|>+<\delta \phi^2_\bot> -v^2)
\label{4}
\end{equation}

\noindent  which clearly differs from the zero temperature one in
presence of fluctuations. By  varying  the  fluctuations,  chiral
symmetry  can  be  restored  or  spontaneously broken. It is also
evident that the evolution of the mean field  critically  depends
on  the initial values of the fluctuations. When $\delta^2 \equiv
(3<\delta \phi^2_\|>-<\delta \phi^2_\bot>)/6$ is large enough the
chiral symmetry is approximately (as H$\neq$ 0) restored. If  the
explicit  chiral  symmetry  breaking term is neglected, the phase
transition   takes   place   at   the    critical    fluctuations
$\delta^2_c\equiv   v^2/6$.  For  $\delta^2  <  \delta^2_c$,  the
effective    potential    takes    its    minimum    value     at
$<\phi>=(\sigma_e,0)$,  where  $\sigma_e$  depends on $\delta^2$.
When the mean fields are displaced from this equilibrium point to
the central  lump  of  the  Mexican  hat  ($<\phi>\sim  0$),  the
effective mass square

\begin{equation}
m_{eff}^2=\lambda(v^2-<\phi>^2-3<\delta \phi^2_\|>-<\delta
\phi^2_\bot>) \label{5}
\end{equation}

\noindent will become negative and DCC can form. Since the domain
size  is  directly  related  to  the time scale, during which the
effective mass remains  negative,  it  strongly  depends  on  the
initial  condition  of  the system. By varying the $<\phi_i>$ and
$\delta^2$, quench or annealing like  initial  condition  can  be
obtained  \cite{as94}.  In  an  important  paper  Asakawa  et  al
\cite{as94}   studied   eq.\ref{1}   with    initial    condition
corresponding to quench and annealing. They found that domains of
disoriented  chiral  condensate  with  4-5  fm  in  size can from
through a quench. Annealing on the otherhand,  leads  to  smaller
sized domains.

In  the  present paper, our interest is to investigate the effect
of friction on DCC domain  formation.  Dissipative  effects  like
friction   damp  the  motion  of  the  fields,  inhibiting  large
oscillations. Moreover, fluctuations-dissipations theorem require
that dissipation be associated with noise. One would then  expect
large  reduction  in  the  DCC  domain  formation  if friction is
present in the system. This naive expectation was  found  to  be
true by Biro and Greiner \cite{bi97}. Using the Langevin equation
for  the linear sigma model, they have investigated the interplay
of friction and  white  noise  on  the  evolution  of  the  order
parameter.  While noise greatly diminishes the possibility of DCC
domain formation, in some orbits, large instabilities can result,
producing DCC  domains.  We  have  also  studied  the  effect  of
friction  on  DCC  domain  formation  using the Langevin equation
\cite{ch99}. There we found that for one-dimensional expansion on
average large DCC domain can not  be  formed.  However,  in  some
particular  orbit large instabilities can occur. This possibility
also reduces with  introduction  of  friction.  However,  if  the
friction is large, the system may be overdamped and then there is
a  possibility  of DCC domain like formation. Present paper is an
extension of the above study, the spatial part,  which  had  been
integrated out in our earlier analysis is being studied here.

Appropriate  coordinates  for heavy ion collisions are the proper
time ($\tau$) and the rapidity (Y). The change can be effected by
the following replacement,

\begin{equation}
\frac{\partial^2}{\partial t^2}- \frac{\partial^2}{\partial z^2}
\rightarrow
\frac{1}{\tau} \frac{\partial}{\partial \tau} \tau
\frac{\partial}{\partial \tau} -
\frac{1}{\tau^2} \frac{\partial^2}{\partial Y^2} \label{6}
\end{equation}

It can be seen from the above equation that with the introduction
of proper time and rapidity, a dissipative term comes into effect
in the equation of motion. To illustrate the role of friction, we
further  introduce  a  dissipative term ($\eta$) in the equation.
However,   fluctuations-dissipation    theorem    require    that
dissipation  be  associated  with fluctuations. We thus include a
white noise term also. To simplify  our  calculation,  we  assume
boost-invariance in the system, Eq.\ref{3} can be written as,

\begin{eqnarray}
\frac{\partial^2 <\phi>}{\partial \tau^2} +(\frac{1}{\tau}+\eta)
\frac{\partial <\phi>}{\partial \tau} =&&
\frac{\partial^2 <\phi>}{\partial x^2} +
\frac{\partial^2 <\phi>}{\partial y^2} +
\lambda[v^2-<\phi>^2
-T^2/2] <\phi> \nonumber \\
&&+H n_\sigma +\zeta(\tau,x,y)
\label{7}
\end{eqnarray}

\noindent    where    $\eta$    is   the   friction  coefficient.
$\zeta(\tau,x,y)$ is a  Gaussian  noise  of  temperature  T  with
correlations,

\begin{mathletters}
\begin{eqnarray}
<\zeta(\tau,x,y)> =&&0\\
<\zeta_a(\tau_1,x_1,y_1)\zeta_b(\tau_2,x_2,y_2)> =&& 2 \eta T
\delta(\tau_1-\tau_2) \delta(x_1-x_2) \delta(y_1-y_2) \delta_{ab}
\end{eqnarray}
\end{mathletters}

It  may  also be noted that we have replaced the fluctuation term
$3<\delta \phi^2_\|>-<\delta  \phi^2_\bot>$  by  its  counterpart
($T^2/2$)    in    the    finite    temperature    field   theory
\cite{ga94,ra97}. The equation of motion of  fields  then  depend
sensitively  on  the  initial  temperature  of the system. In the
following,    we    assume    initial    temperature    to     be
$T_c=\sqrt{2f_\pi^2-2m_\pi^2/\lambda}$=123  MeV  at  the  initial
time $\tau_i$= 1 fm. The cooling of the system  is  described  by
the  following  equation  appropriate  for  scaling  expansion in
2-dimension,

\begin{equation}
\frac{\dot{T}}{T} + \frac{2}{3\tau}=0 \label{8}
\end{equation}

The  friction  coefficient  was  assumed  to  be  $\eta=\eta_\pi +
\eta_\sigma$,  where   $\eta_{\pi,\sigma}$   are   the   friction
coefficient  of  the  pion  and  the sigma fields. They have been
calculated by Rischke \cite{ri98},

\begin{mathletters}
\begin{eqnarray}
\eta_\pi =&& (\frac{4 \lambda f_\pi}{N})^2 \frac{m_\sigma^2}{4  \pi
m_\pi^3} \sqrt{1-\frac{4 m_\pi^2}{m_\sigma^2}}
\frac{1-exp(-m_\pi/T)}{1-exp(-m_\sigma^2/2 m_\pi T)}
\frac{1}{exp(m_\sigma^2-2m_\pi^2)/2m_piT)-1}\\
\eta_\sigma   =&&  (\frac{4  \lambda  f_\pi}{N})^2  \frac{N-1}{8\pi
m_sigma}
\sqrt{1-\frac{4 m_\pi^2}{m_\sigma^2}} \coth \frac{m_\sigma}{4T}
\end{eqnarray}
\end{mathletters}

We  solve  the set of partial differential equations \ref{7} with
the quenched initial condition. Accordingly  the  initial  fields
are  randomly  distributed  to a Gaussian form with the following
parameters,

\begin{mathletters}
\begin{eqnarray}
<\sigma>=&&(1-f(r))f_\pi \\
<\pi_i>=&&0 \\
<\sigma^2>-<\sigma>^2 = <\pi_i^2>-<\pi_i>^2=   && v^2/4 f(r)\\
< \dot{\sigma}>=&& <\dot{\pi_i}>=0\\
<\dot{\sigma}^2>=<\dot{\pi}>^2=&& v^2
\end{eqnarray}
\end{mathletters}

The interpolation function

\begin{equation}
f(r)=[1+exp(r-r_0)/\Gamma)]^{-1}
\end{equation}

\noindent separates the central region where  the  initial  field
configuration  is  different from their vacuum expectation value.
We use $r_0$=5 fm and $\Gamma$=0.5 fm.

The  equation of motion was solved for 500 trajectories. For each
trajectories, at each space-time we compute  the  effective  mass
$m^2_{eff}$.  The phenomenon of long wavelength DCC amplification
will occur whenever the effective mass squared  is  negative.  To
single out the trajectories for which maximum instabilities occur
we calculate the following quantity,

\begin{equation}
G=\int  |m_{eff}| \Theta(-m_{eff}^2) \tau d\tau dx dy
\end{equation}

This can be a measure of instability in a particular evolution. We
call  it amplification factor. This is an important parameter, as
it directly relates to the size of DCC domains. In fig.1, we have
shown the distribution of the amplification factor G for the  500
trajectories. The distribution is more or less Gaussian like. All
the    trajectories  shows  appreciable  instability.  The
maximum and the minimum instabilities differing by  $\sim$  10\%.
Unlike  1-dimensional  expansion,  DCC  domain  formation  is   a
distinct possibility in 2-dimensional expansion scenario.  It  is
interesting  to  note  that presence of friction and noise do not
change the scenario to a great extent.

In  fig.2, we show the contour plot of $\pi_2$ field for the most
unstable orbit. Initially,  at  1  fm,  the  field  are  randomly
distributed. There is no domain like structure. After a few fermi
of  evolution,  correlation starts to build up, and a domain like
structure can be clearly seen. The domain like structure is  most
distinguished  at  5 fm, where we see clear two domain formation.
It may be noted that the two domains  have  different  sign.  The
domain like structure persist at 7 fm also. The result is a clear
indication of DCC like phenomena, even in presence of dissipation
and noise.

In  fig.3,  we  have  shown the time evolution of the correlation
function, defined as,

\begin{equation}
C(r,\tau)=\frac{\Sigma_{ij} \pi(i) \times \pi(j)}
               {\Sigma_{ij} |\pi(i)| \times |\pi(j)|}
\end{equation}

\noindent  where  the sum is taken over those grid points $i$ and
$j$ such that the distance between points is $r$. At 1 fm,  there
is  no correlation beyond the lattice spacing (.25). A long range
correlation emerges at 3 fm.  The  correlation  length  increases
as the evolution time increases and at 5 fm it is as large  as  5
fm, corroborating the pictorial result of fig.2

To  summarize, we have investigated the effect of friction on the
possible DCC domain formation. In the equation of motion for  the
linear  sigma  model  fields,  we  include a friction and a white
noise term and solve it assuming boost invariance. The noise term
is required to be  consistent  with  the  fluctuation-dissipation
theorem.  Initial  field configuration was assumed to be Gaussian
random with quenched condition ($<\phi>=\dot{<\phi>}=0$). At each
space-time  $m^2_{eff}$  was  calculated.  Phenomenon   of   long
wavelength  amplification  occurs  when  it becomes negative. For
each trajectories, we  calculate  the  amplification  factor,  as
defined  in  eq.().  Amplification  factor gives an indication of
instability  in  the  system.  It  was  seen  that  for  the  500
trajectories, the maximum and minimum instabilities do not differ
much,  indicating  possibility of domain like structure formation
in all the events. Clear domain like formation  is  seen  in  the
event   with   maximum  instability.  For  initial  random  field
distribution, domain like structure emerges after a few fermi  of
evolution.  Most  distinct domain like structure is seen at 5 fm.
There domain of 5 fm size is seen to  be  formed.  We  have  also
calculated  the  correlation function for the pion field. Similar
result is also obtained there. Initially correlation donot extend
beyond  .25  fm,  (the  lattice  spacing).  After  a  few  fm  of
evolution,  long  range  correlation starts to build up. At 5 fm,
correlation length of  5  fm  is  obtained.  The  result  clearly
indicate that the DCC domain formation is a distinct possibility,
even in presence of friction and noise.

\begin{figure}
\centerline{\psfig{figure=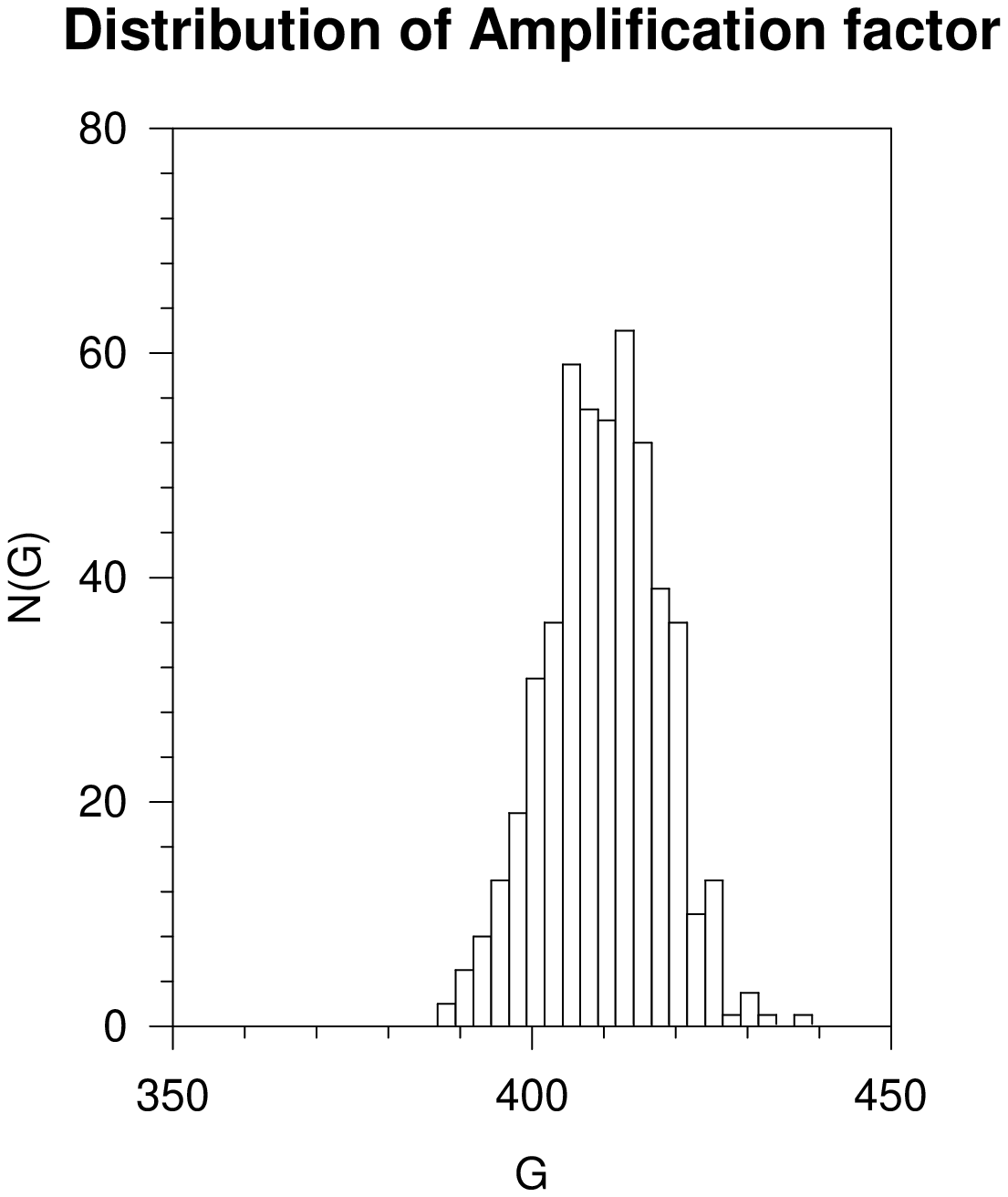,height=15cm,width=10cm}}
\vspace{-5cm}
\caption{Distribution  of  the amplification factor G for the 500
trajectories calculated.}
\end{figure}
\begin{figure}
\centerline{\psfig{figure=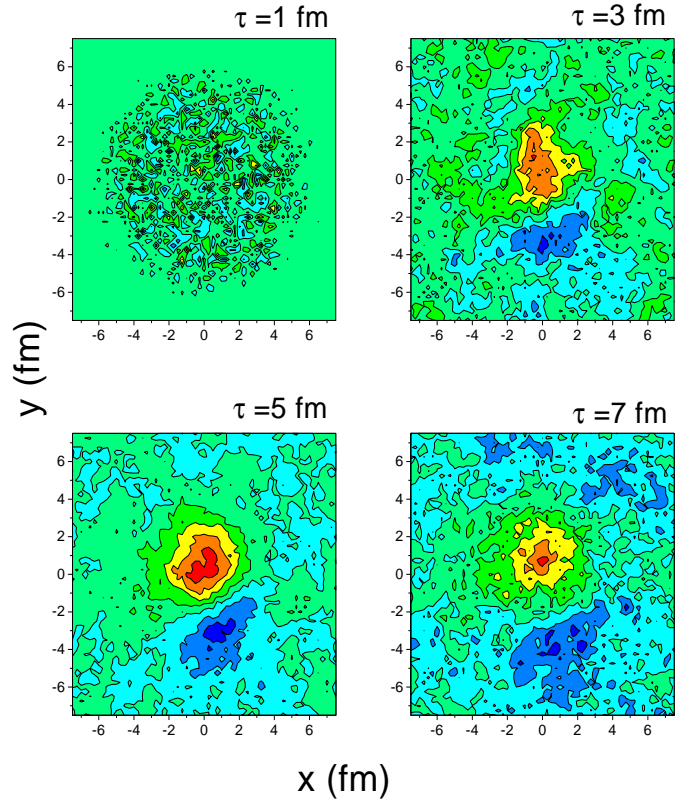,height=15cm,width=10cm}}
\vspace{-1cm}
\caption{Contour  plot of evolution  of  $\pi_2$  field  for  the
event  with  maximum  instability.  At  5  fm,  two  domain  like
structure can be seen clearly. They have opposite sign.}
\end{figure}
\begin{figure}
\centerline{\psfig{figure=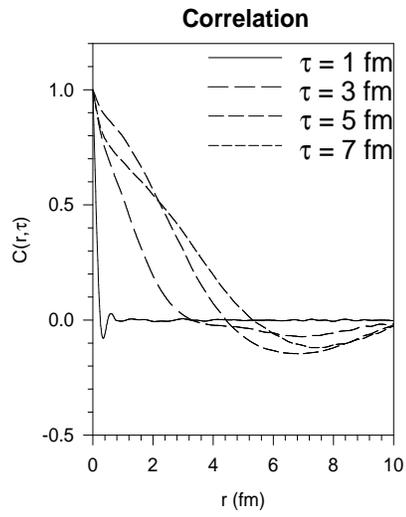,height=15cm,width=10cm}}
\vspace{-5cm}
\caption{Evolution of the correlation function. Initially at 1 fm,
there is no correlation beyond .25 fm, the lattice spacing. Large
correlation devolop at later times.}
\end{figure}
\end{document}